\documentclass{article}

\usepackage{arxiv}

\usepackage[utf8]{inputenc} 
\usepackage[T1]{fontenc}  
\usepackage{hyperref}    
\usepackage{url}      
\usepackage{booktabs}    
\usepackage{amsfonts}    
\usepackage{nicefrac}    
\usepackage{microtype}   
\usepackage{lipsum}		
\usepackage{graphicx}
\usepackage{doi}
\usepackage{mathtools}
\usepackage{amsmath}

\usepackage[numbers,sort&compress]{natbib}
\usepackage{hyperref}

\usepackage[table,xcdraw]{xcolor}

\usepackage{colortbl}

\usepackage{float}

\usepackage{enumitem} 
\usepackage{array}  

\usepackage{graphicx,subcaption} 

\DeclareUnicodeCharacter{2212}{-}

\newcommand{\openr}{\hbox{${\rm I\kern-.2em R}$}}

\title{HAL-Based Plug-in Estimation with Pointwise Asymptotic Normality of the Causal Dose-Response Curve}

\author{\href{https://orcid.org/0000-0002-1520-8387}{\includegraphics[scale=0.06]{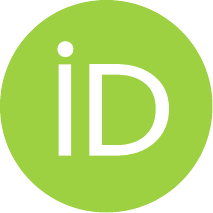}\hspace{1mm}Junming Shi}\\
	Department of Biostatistics\\
	University of California, Berkeley\\
	Berkeley, CA 94704; \\
        Department of Radiation Oncology,\\
        Stanford University,\\
        Stanford, CA 94035\\
  \texttt{junmings@stanford.edu} \\
	\And
	\href{https://orcid.org/0009-0008-9106-1672}{\includegraphics[scale=0.06]{orcid.pdf}\hspace{1mm}Wenxin Zhang} \\
	Department of Biostatistics\\
	University of California, Berkeley\\
	Berkeley, CA 94704 \\
	\texttt{hubbard@berkeley.edu} \\
  \And
	\href{https://orcid.org/0000-0002-3769-0127}{\includegraphics[scale=0.06]{orcid.pdf}\hspace{1mm}Alan E. Hubbard} \\
	Department of Biostatistics\\
	University of California, Berkeley\\
	Berkeley, CA 94704 \\
	\texttt{hubbard@berkeley.edu} \\
		\And
	\href{https://orcid.org/0000-0003-4853-6130}{\includegraphics[scale=0.06]{orcid.pdf}\hspace{1mm}Mark J. van~der~Laan}\\
	Department of Biostatistics\\
	University of California, Berkeley\\
	Berkeley, CA 94704 \\
	\texttt{laan@berkeley.edu}
}

\renewcommand{\shorttitle}{\textit{HAL-based plugin estimation for Dose-Response Curve}}

\begin{document}
\maketitle

\begin{abstract}
Estimating and obtaining reliable inference for the marginally adjusted causal dose-response curve for continuous treatments without relying on parametric assumptions is a well-known statistical challenge. Parametric models risk introducing significant bias through model misspecification, compromising the accurate representation of the underlying data and dose-response relationship. On the other hand, nonparametric models face difficulties as the dose-response curve is not pathwise differentiable, preventing consistent estimation at standard rates. The Highly Adaptive Lasso (HAL) maximum likelihood estimator offers a promising approach to this issue. In this paper, we introduce a HAL-based plug-in estimator for the causal dose-response curve, bridge theoretical development and empirical application, and assess its empirical performance against other estimators. This work emphasizes not just theoretical proofs, but also demonstrates their application through comprehensive simulations, thereby filling an essential gap between theory and practice. Our comprehensive simulations demonstrate that the HAL-based estimator achieves pointwise asymptotic normality with valid inference and consistently outperforms existing approaches for estimating the causal dose-response curve.

\end{abstract}


\newpage

A marginally adjusted dose-response curve, often referred to as a "causal" dose-response curve, provides a graphical representation of the population-level probability of a health outcome following a uniform application of a specific dose or, more generally, a dosage rule applied to the target population. Estimation of such a curve where the dose is ordinal or continuous requires accounting for potential confounders, missingness, and other typical complications in data-generating processes. Efficient estimators with valid inference are crucial for a wide range of scientific applications, including epidemiology, medical research, social sciences, and many other fields. For example, in a clinical trial investigating the impact of a continuous dose of drugs on blood pressure reduction, the dose-response curve can provide valuable insights into the specific reduction or increase in blood pressure attributed to each incremental change in dosage, accounting for confounding factors such as demographic and baseline clinical characteristics.

To estimate the causal dose-response curve, we need to define a parameter under the counterfactual or potential outcome framework in which the outcome would be observed for subjects in the target population if all of them received a specific treatment level (van~der~Laan and Petersen, 2028; Petersen et al., 2006; Rubin, 2004). This framework, widely used in causal inference, has predominantly focused on binary interventions (van~der~Laan et al., 2011; Rubin, 2006; van~der~Laan and Robins, 2003; Abadie and Imbens, 2006; Angrist and Imbens, 1995; Robins, 1986; Rubin, 1978). Common approaches for estimating causal effects, such as the average treatment effect (ATE), include matching (Abadie and Imbens, 2011), estimating equations (e.g., inverse probability of treatment weighting (IPTW) (Austin and Stuart, 2015) and augmented IPTW (A-IPTW) (Luque-Fernandez et al., 2018)), and substitution estimators (e.g., G-computation (Robins, 1986) and targeted maximum likelihood estimation (TMLE) (van~der~Laan et al., 2011)). Furthermore, highly adaptive lasso (HAL) (van~der~Laan, 2017) and undersmoothed HAL (van~der~Laan et al., 2022) have been developed to provide reliable inference for the effects of binary treatment. Although significant progress has been made in the binary treatment context, many causal inference questions inherently involve continuous or ordinal interventions, such as doses in clinical treatments, for which consistent and efficient estimators remain scarce.

The primary challenge in estimating the causal dose-response curve for continuous treatments in a nonparametric model is that the parameter is not pathwise differentiable (Bickel et al., 1993). Consequently, $\sqrt{n}$ estimators are not applicable, leading to reliance on estimators that depend on estimating densities rather than a smooth function of this density.

Considerable efforts have been dedicated to crafting estimators for the causal dose-response curve. The Marginal Structural Model (MSM) framework, originally introduced by Robins et al. (2000), is known for defining causal dose-response curves in parametric models. However, these methods rely on specifying parametric models of the dose-reponse curve (MSM) and thus is subject to incorrect model specification.  To address potential model misspecification, enhancements to MSM have been proposed using outcome prediction models, data-adaptive estimation of MSM, and projecting the true causal dose-response curve onto a functional space defined by MSM (Robins and Greenland, 2000; Haight et al., 2010; Neugebauer and van der Laan, 2007). Despite these advances, challenges related to robustness remain. Alternative estimators based on plug-in methods for conditional outcome models have been proposed (Hirano and Imbens, 2004; Callaway et al., 2021; Schwab et al., 2020; Díaz and van~der~Laan, 2013), but these often struggle with dimensionality issues, restrictive assumptions, and the complexity of valid inference. Techniques such as cross-validated TMLE (Chapter 25 in van der Laan and Rose, 2018) and nonparametric causal inference (Kennedy et al., 2017) achieve asymptotic normality, facilitating inference, but are constrained by their reliance on smoothness assumptions. Additionally, their reliance on kernel smoothing introduces complexities in selecting the tuning parameter, particularly given the suboptimal performance of global cross-validation selectors for smoothness and bandwidth selection (Díaz et al., 2023).

In this article, we present and enhance the highly adaptive lasso (HAL) based plug-in estimator for the marginally adjusted causal dose-response, with a focus on bridging the gap between theory and empirical application. HAL-MLE is a nonparametric minimum-loss estimator capable of efficiently estimating pathwise differentiable parameters. Importantly, it has been theoretically proven to perform well with pointwise asymptotic normality in settings where the parameter is non-pathwise differentiable (van~der~Laan, 2023), such as with continuous treatments. Specifically, the HAL-based plug-in estimator converges to the true curve at a dimension-free rate up to log n factors (van~der~Laan, 2023). This work aims to implement the HAL-based estimator and validate its theoretical properties through empirical analysis.

Through simulation studies, we assess the performance of this HAL-based estimator under various realistic conditions, showing that it provides valid inference with robust confidence intervals for the causal dose-response curve. These simulations confirm that the theoretical properties of the HAL estimator are effectively translated into practice, offering a thorough empirical validation of its utility for continuous dose-response estimation. Our contribution extends the utility of HAL to continuous dose-response problems, which have traditionally been difficult to estimate efficiently due to nonpathwise differentiability. We present the theoretical foundations of the HAL-based plug-in estimator and its associated delta-method-based inference that we implemented, followed by a series of simulations demonstrating its consistent performance and reliability.

In the sections that follow, we define the target parameter and present the HAL-based plug-in estimator with confidence intervals based on the delta method. Then, we present simulation results that highlight the estimator's asymptotic normality, nominal coverage, and overall reliability for estimating the marginally adjusted dose-response curve in continuous treatment settings. 

\section{Method}
\label{s:method}

\subsection{Target Parameter and Statistical Model}

Consider an observational study or a randomized experiment with $n$ independent and identically distributed observations of a continuous exposure or treatment variable $A$, an outcome $Y$, and covariates $W$, represented as $O = (W, A, Y)$. $O$ follows a true data-generating distribution (DGD) $P_0 \in {\cal M}^{NP}$, where $\mathcal{M}^{NP}$ is a nonparametric model that places minimal assumptions on the data-generating process, thus allowing for a more flexible and realistic representation of complex relationships among variables. The goal is to estimate the marginal causal dose-response curve, which captures the causal relationship between treatment and outcome. The target parameter $\Psi_P(a)$ is defined as:
\[
    \Psi_P(a) = \mathbb{E}_P[Y(a)] = \mathbb{E}_P[\mathbb{E}_P[Y|W,A=a]],
\]
where $\Psi_P(a)$ represents the expected outcome if all subjects were treated with $A = a$. The parameter $\Psi_0 = \Psi(P_0)$ is the value of this estimand for the true data-generating distribution $P_0$. 

When $Y$ is continuous, we focus on estimating the conditional mean $Q_P(W,A) := \mathbb{E}_P[Y|W,A]$, such that $\Psi_P(a) = \int_w Q(w,a) dP(w)$. This leads to the plug-in estimator $\hat \Psi(a) = \frac{1}{n} \sum_{i=1}^n \hat{Q}(W_i, a)$, where $\hat{Q}$ is a nonparametric estimator of the conditional mean, such as the Highly Adaptive Lasso (HAL).  When $Y$ is binary, we parameterize $E_P(Y\mid W,A)=1/(1+\exp(-Q_P(W,A))$ so that $Q_P$ now plays the role of $\mbox{Logit}E_P(Y\mid W,A)$ as the conditional probability. 

Estimating this target parameter is challenging, since there is no efficient estimation method with reliable inference. Therefore, we adopt a flexible and realistic statistical model based on càdlàg functions with finite sectional variation norm. This approach allows for robust estimation without making unrealistic assumptions.

\subsection{Flexible and Realistic Statistical Model: Càdlàg Functions}

To model the relationship between $A$, $Y$, and $W$ without relying on strong parametric assumptions, we use the class of multivariate càdlàg (right-continuous with left limits) functions with finite sectional variation norms (FSVN) as the foundation of the statistical model. This class includes not only continuous functions but also functions with jumps and other non-linear features, providing high flexibility for approximating complex relationships in real-world data while avoiding strict smoothness assumptions.

For a $d$-variate càdlàg function $Q: [0,1]^d \rightarrow \mathbb{R}$, the function $Q(x)$ can be represented as:
\[
    Q(x) = Q(0) + \sum_{S \subset \{1,\ldots,d\}} \int_{(0(S), x(S)]} Q_S(du),
\]
where for each subset $S \subset \{1, \dots, d\}$, $x(S)$ denotes the vector of coordinates in $S$, and $Q_S$ is the $S$-specific section of $Q$. The measure $Q_S(du)$ represents the measure on $(0(S), x(S)]$ generated by the càdlàg function $Q_S$. The sectional variation norm, which measures the total variation of the function over all sections, is defined as:
\[
||Q||^*_v = |Q(0)| + \sum_{S \subset \{1, \dots, d\}} \int_{(0(S), 1(S)]} |dQ_S(u)|.
\]

Càdlàg functions form a Donsker class, ensuring that empirical processes indexed by this class converge to Gaussian processes. This convergence provides desirable statistical properties, such as optimal convergence rates and asymptotic normality (Van der Vaart et al., 1996). These properties make càdlàg functions an attractive choice for constructing estimators in complex, high-dimensional data settings. In particular, Highly Adaptive Lasso (HAL) leverages this flexibility to estimate the target function $Q$ robustly, while achieving optimal convergence rates under weak assumptions.

\subsection{Highly Adaptive Lasso (HAL) and Representation of $Q$: From Statistical Model to Practical Estimator}

Building on the flexibility of càdlàg functions, the Highly Adaptive Lasso (HAL) is a nonparametric minimum-loss estimator that can flexibly estimate complex, high-dimensional target parameters. In its most nonparametric form, HAL only assumes that the true target function belongs to the class of $d$-variate real-valued càdlàg functions with bounded sectional variation norms. This allows HAL to fit a wide range of functions, including those with discontinuities and nonlinear relationships, making it a powerful tool for data-driven estimation without imposing unrealistic parametric assumptions (van~der~Laan, 2015).

At the heart of HAL is the minimization of an empirical loss function, subject to a constraint on the function's sectional variation norm, which effectively controls the complexity of the model. This combination allows HAL to achieve near-optimal convergence rates and provides a robust estimation of complex relationships between variables.

\subsubsection{Representation of $Q$ in HAL: Zero-Order Approximation}

For a $d$-dimensional càdlàg function $Q: [0,1]^d \rightarrow \mathbb{R}$, HAL approximates $Q$ using a finite linear combination of indicator (zero-order spline) basis functions. This can be represented as:
\[
Q_M(x) = Q(0) + \sum_{S \subset \{1,\ldots,d\}} \sum_{j} I(s_j(S) \leq x(S)) dQ_{M,S,j},
\]
where $I(s_j(S) \leq x(S))$ is an indicator function that captures whether the observation $x$ falls within a certain region of the domain defined by support points $s_j$, and $dQ_{M,S,j}$ are the coefficients. The sectional variation norm in this case is:
\[
\|Q_M\|_v^* = |Q(0)| + \sum_{S \subset \{1,\ldots,d\}} \sum_{j} |dQ_{M,S,j}|,
\]
which measures the total variation of the function across all sections, controlling the complexity and smoothness of the estimator.

HAL solves for the estimator by minimizing the empirical loss:
\[
Q(P) = \arg\min_{Q \in D^{(0)}_M([0,1]^d)} P_n L(Q),
\]
where $P_n L(Q)$ is the empirical loss and $D^{(0)}_M([0,1]^d)$ is the space of zero-order càdlàg functions with a bounded sectional variation norm $M$. This formulation ensures that the estimated function adheres to the constraint $||Q||^*_v < M$, promoting smoothness and preventing overfitting.

\subsubsection{Approximation in Practice: Basis Expansion with Knot Points}

In practice, HAL approximates $Q$ by constructing a large set of zero-order spline basis functions using knot points from the observed data. Given the input vector $X=(W,A)$, the basis functions are derived from the support points $\{X_i(S):1,\dots,n, S \subset \{1, \dots,d\}\}$. The resulting approximation is expressed as follows: 
\[
\begin{split}
Q_{\beta}(x) & = \beta_0 + \sum_{s_1 \subset \{1,2,...,d\}}\sum_{j=1}^{n} \beta_{s_1,j} \phi_{u_s,j}(x) = \boldsymbol{\Phi}^0(x) \boldsymbol{\beta} \\
\hspace{0.8cm}& \mbox{with}\hspace{0.8cm} 
||{\beta}||_1 = |\beta_0| + \sum_{s_1\subset \{1,2,...,d\}}\sum_{j=1}^{n} |\beta_{s,j}| < M.
\end{split}
\]
This approximation is a linear combination of zero-order basis functions $x \rightarrow \phi_{u_s,j}(x) = I(x_{u_s, j} \leq x_{s})$, defined by support points $x_{u_s,j}$ from observations, along with their corresponding coefficients $\beta_{s,j}$. 

\subsubsection{Higher-Order HAL: First-Order and Higher}

HAL can be extended to estimate smoother functions by incorporating higher-order splines. The first-order HAL assumes that the true function $Q$ belongs to the class $D^{(1)}([0,1]^d)$, which consists of functions whose sections are absolutely continuous with respect to the Lebesgue measure. The derivative of $Q_S$ with respect to $x(S)$, denoted as $Q^{(1)}_S \equiv dQ_S/dx(S)$, is itself a càdlàg function with bounded sectional variation.

The first-order HAL approximates $Q$ using first-order spline basis functions, which allow smoother representations of the target function. The general form of the first-order HAL approximation is:
\[
Q_\beta(x) = \beta_0 + \sum_{S \subset \{1,\ldots,d\}} \sum_{j} \beta_{S,j} \phi^1_{s_j(S)}(x(S)),
\]
where $\phi^1_{s_j(S)}(x(S))$ represents the first-order spline functions:
\[
\phi^1_{s_j(S)}(x(S)) = I(x_j \geq s_j)(x_j - s_j),
\]
with $\beta_0$ and $\beta_{S,j}$ being the model coefficients. The $L_1$-norm of the coefficients corresponds to the first-order sectional variation norm $\|Q\|_{v,1}^*$. This norm controls the smoothness of the function while allowing for greater flexibility than the zero-order HAL model.

For example, in the case of $d=1$, the first-order spline function is:
\[
\phi^1_u(x) = (x - u) \mathbb{I}(x \geq u),
\]
and for $d=2$, it becomes:
\[
\phi^1_u(x_1, x_2) = (x_1 - u_1)(x_2 - u_2) \mathbb{I}(x_1 \geq u_1, x_2 \geq u_2).
\]
These spline functions enable the first-order HAL model to capture interactions between variables and smooth relationships across dimensions.

Extending beyond the first order, HAL can be generalized to accommodate $k$-th order smoothness classes $D^{(k)}([0,1]^d)$, where $k = 1,2,\ldots$. Functions in $D^{(k)}([0,1]^d)$ can be approximated by a finite linear combination of $k$-th order splines. The $L_1$-norm corresponds with the sum over all $k$-th-order derivatives of its sectional variation norms, referred to as the $k$-th order sectional variation norm. See works by van~der~Laan (2023) for detailed theoretical development.

\subsection{Optimizing HAL fit of Dose-Response Curve}

\subsubsection{Sectional Variation Norm Bound (Penalty) Selector}

The HAL estimator is constructed to minimize the empirical loss while selecting an appropriate bound for the sectional variation norm. The choice of the $L_1$-norm bound, $M$, is crucial for balancing bias and variance in the estimation. Cross-validation is used to select the optimal $M$. However, when estimating the dose-response curve, this often results in oversmooth estimators, biased for the functional of interest. Therefore, an undersmoothing strategy is employed, in which the bound is increased beyond the cross-validated value to reduce bias (van~der~Laan et al., 2022).

In undersmoothed HAL, the $L_1$-norm bound $M$ is selected to be sufficiently large, larger than the value determined through cross-validated, to ensure the efficient influence curve score equation is solved at the level $\hat \sigma_n/(n^{1/2}\log n)$, where $\hat \sigma_n/n^{1/2}$ is the standard error estimate based on the sample variance of the efficient influence curve. This approach is analogous to selecting a smaller penalty, $\lambda$, in GLMNET, compared to the cross-validated value,  with the goal of solving the efficient influence curve score equations.

Undersmoothing ensures that the linear span of the basis functions captures sufficient complexity to approximate the efficient influence curve of the target parameter, which lies within the linear span of the score functions. This reduces bias relative to sampling variance and improves the fit of the dose-response curve. The HAL estimator optimizes the fit of the càdlàg function $Q$ by achieving dimension-free convergence rates, up to a logarithmic factor.

\subsubsection{Smoothness Order Selector}

In practical scenarios, the exact shape of the dose-response curve is often unknown, posing challenges in determining the optimal level of smoothness and other crucial hyperparameters for the Highly Adaptive Lasso (HAL) fit, which depend on the curve’s characteristics. To address these challenges, the Smoothness-Adaptive HAL, introduced by van~der~Laan (2023), leverages the discrete Super Learner to identify the appropriate smoothness order and other relevant hyperparameters, such as the number of knots, within a cross-validation framework. This approach offers a robust data-driven method for selecting the hyperparameter that yields the best empirical performance in modeling the causal dose-response curve.

\subsubsection{Dose-Response Curve Estimation with HAL}

After determining the optimal hyperparameter configuration, undersmoothing is applied to further refine the HAL fit. This step ensures that the bias is of second order relative to the sample variance, balancing the precision and robustness of the estimated dose-response curve.

The dose-response curve $\Psi_P(a)$ is then estimated via undersmoothed HAL as:
\[
\hat{\Psi}(a) = \frac{1}{n} \sum_{i=1}^n \hat{Q}(W_i, a),
\]
where $\hat{Q}$ is the HAL-based estimator of the conditional mean function $Q_P$. This plug-in estimator aggregates predictions from the HAL model across all observed covariate values $W_i$, providing a nonparametric estimate of the causal dose-response relationship.

\subsection{Inference: Pointwise Asymptotic Normality of Delta-Method-Based Estimators and Confidence Intervals}

The HAL estimator has been shown to be asymptotically normally distributed (van~der~Laan, 2023). This property allows us to conduct inference using the delta method, providing a mechanism for constructing confidence intervals (CIs) for the estimated dose-response curve. The variance of $Q_{\beta_n}(x)$, where $Q_{\beta_n}$ is the HAL-based estimator, is estimated by the influence curve of the coefficients $\beta_n$. 

For a continuous outcome, HAL solves the score equation:
\[
P_n \frac{d}{d\beta}L(Q_{\beta}) = 0,
\]
where $\beta_n = \arg\min_{\beta} P_n L(Q_{\beta})$. Using the delta method, the plug-in estimator $\hat{\Psi}(a)$ is asymptotically linear, with the influence curve:
\[
IC_{\Psi(a),P}(O_i) = \mathbb{E}_W[\Phi(a, W)^T IC_{\beta, P}(O_i)],
\]
where $\Phi(a, W)$ is the basis vector for treatment $A = a$ and covariates $W$, and $IC_{\beta, P}(O_i) = \mathbb{E}_P[\Phi_i^T\Phi_i]^{-1}\Phi_i^T(Y_i-\Phi_i^T\beta)$ is the influence curve of the coefficients. The variance of the estimator $\hat{\Psi}_n(a)$ is estimated by computing the empirical variance of the influence curve:
\[
\hat{\sigma}^2_n(a) = \frac{1}{n} \hat{\text{Var}}(IC_{\Psi(a), P}(O_i)),
\] where $\hat{\text{Var}}$ denotes the sample variance computed from the estimated influence curve evaluated on the observed data. Because $\hat{\sigma}_n^2(a)$ estimates the variance of $\hat{\Psi}_n(a)$, it can be directly used to construct confidence intervals. 

The corresponding 95\% confidence interval for the dose-response curve is then given by:
\[
\hat{\Psi}_n(a) \pm z_{1-\frac{\alpha}{2}}\hat{\sigma}_n(a),
\]
where $z_{1-\frac{\alpha}{2}}$ is the $1-\frac{\alpha}{2}$ quantile of the standard normal distribution ($\alpha = 0.05$).

These working model-based confidence intervals are asymptotically reliable according to the theoretical results in van~der~Laan (2023). While they do not account for potential bias, they provide inference for the projection $Q_{\beta_n}$. However, with appropriate undersmoothing, this bias becomes asymptotically negligible relative to the standard error. 

\subsection{Simulation settings}
To assess the effectiveness of HAL-based estimations in determining the marginal dose-response curve $\mathbb{E} Y(a)$, Monte Carlo simulations were conducted for each of four different DGDs, each with qualitatively different marginal dose-response curves, $\Psi(a)(P_0)$. These curves varied in complexity, continuity, and the presence of discontinuity. 

\begin{figure}[h]
 \begin{center}
     \includegraphics[width=0.6\textwidth]{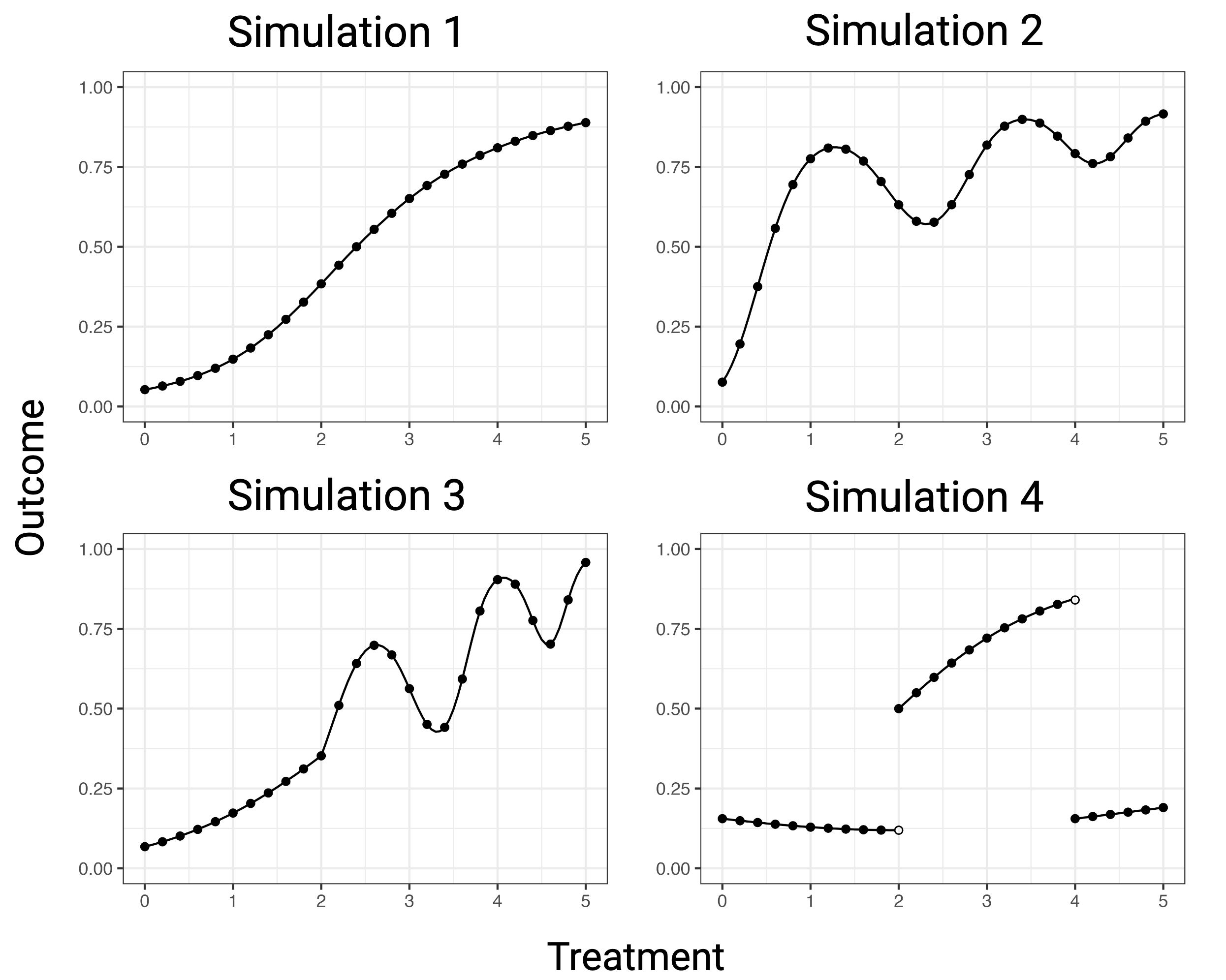}
 \end{center}
 \caption{\textbf{True causal dose-response curves across four simulations.} The ground-truth dose-response relationships exhibit increasing complexity, ranging from monotonic trends (Simulation 1) to highly non-linear and discontinuous patterns (Simulations 2–4).}
 \label{fig:true_curves}
\end{figure}

Figure~\ref{fig:true_curves} illustrates the true marginal dose-response curve $\Psi(P_0)(a)$ for each simulation. In Simulation 1, the curve is smooth and monotonically increasing, albeit non-linear. Simulation 2 presents a curve with several oscillations. Simulation 3 combines characteristics from simulations 1 and 2: the first half of the curve is smooth, resembling that of simulation 1, while the segment where $a$ lies between 2 and 5 exhibits oscillations. In simulation 4, the curve is characterized by two discontinuities at a = 2 and a = 4, with non-zero values exclusively within this interval.

These four simulations were executed under a consistent setting. For each simulation, $n$ independent and identically distributed (i.i.d.) copies of the data $O=(W,A,Y)$ were simulated. Here, $W$ represents a baseline covariate, $A$ denotes the treatment level, which depends on $W$ and varies continuously between 0 and 5, and $Y$ is the outcome of interest, denoting the occurrence of an event.

In each simulation, we generated 500 datasets with sample sizes of 200, 500, 1000, and 5000, following the data-generating processes outlined in Table ~\ref{tab:DGDs} in Appendix A. These datasets were fitted by six variations of Highly Adaptive Lasso (HAL), with each model differing in terms of smoothness and regularization methods. These included: (1) zero-order smoothness with cross-validation (CV) selector, (2) zero-order smoothness with undersmoothing, (3) first-order smoothness with CV selector, (4) first-order smoothness with undersmoothing, (5) adaptive smoothness with CV selector, and (6) adaptive smoothness with undersmoothing. The zero-order models were configured using the default number of knots from the hal9001 R package (Hejazi et al., 2020), catering to scenarios with indeterminate smoothness orders and knot configurations. In contrast, first-order models were trained with a smaller number of knots, suited to simulated scenarios with lower dimensionality and complexity. The adaptive models dynamically adjusted to the data's complexity, providing increased flexibility in the fitting process. For each simulation setting, oracle standard deviations were calculated across the 500 repetitions, with corresponding confidence intervals derived, to evaluate coverage under consistent estimation of the sampling variability and thus the estimators' asymptotic normality. 

These variants are evaluated based on five metrics from both perspectives of delta-method-based inferences and Monte Carlo oracle performances: coverage rates of 95\% confidence intervals, bias, SE, bias/SE ratio, and mean squared error (MSE).   To estimate coverage, we report both that based on the estimated delta-method-based standard error and on the {\em oracle} standard error, calculated as the standard deviation of the estimator over the repeated simulations.  We did this to separate biases in estimating the standard error from non-normal finite sampling distributions.   To evaluate the selected penalties by CV and global undersmoothing criteria, comparisons were made against a grid of penalties. Furthermore, we compared the performance of globally undersmoothed smoothness adaptive HAL against established models, including the general adaptive model (GAM), polynomial regression model, and the nonparametric causal inference (npcausal) model by Kennedy et al. (2017).

\section{Results}
\label{s:results}

In this section, we present our key findings across four perspectives: (1) Cross-Validation (CV) and the implications of Undersmoothing, (2) Evaluation of Undersmoothing Criteria, (3) Analysis of the Smoothness-Order Adaptive HAL, and (4) Comparative Assessment with other methods. To substantiate our findings, we reference one or two simulations as examples. Detailed results shown in comprehensive plots for all simulations and sample sizes are provided in online supporting materials.

\subsection{CV and Undersmoothing}

\begin{figure}[h]
\begin{center}
    \includegraphics[width=0.8\textwidth]{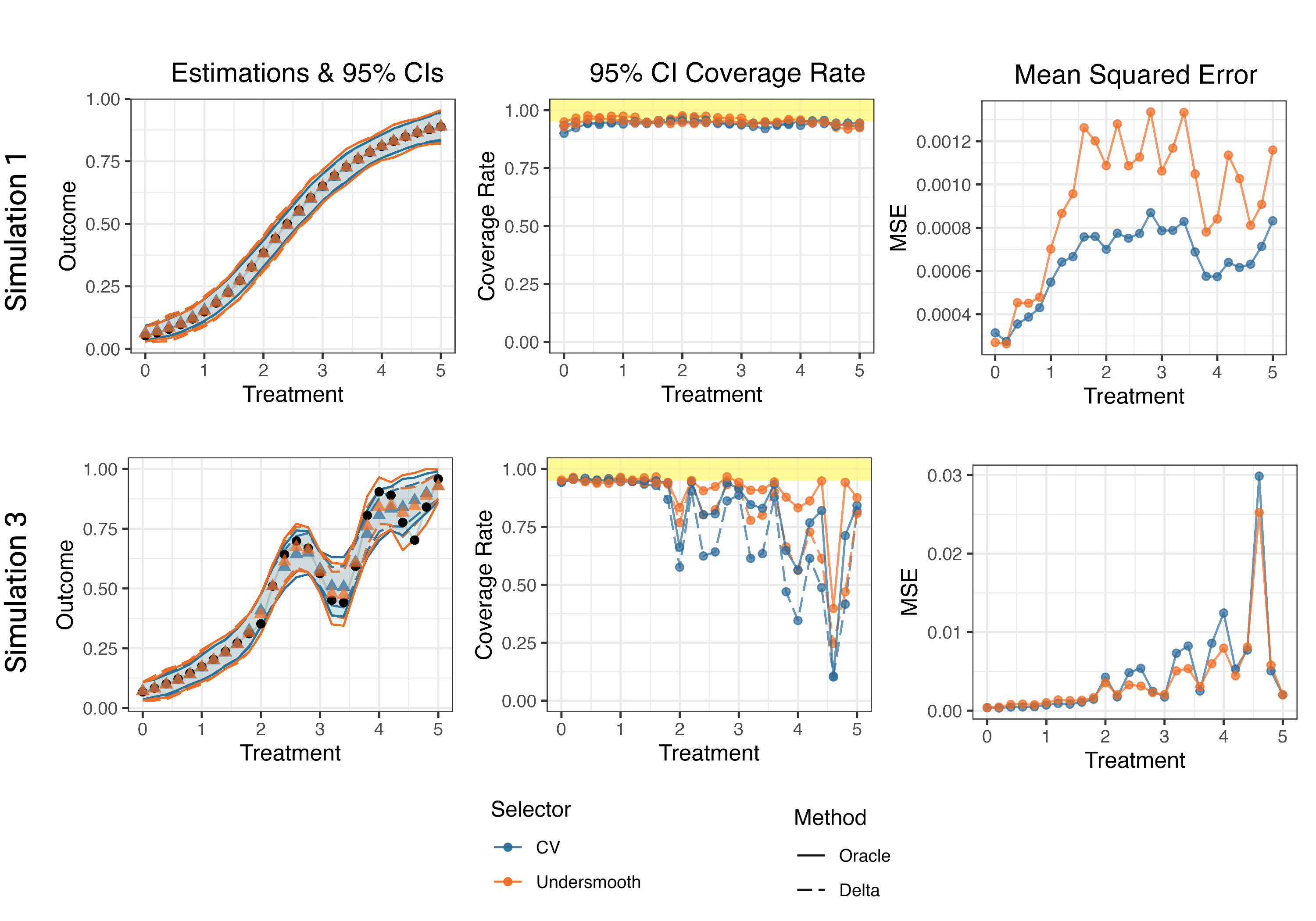} 
\end{center}
\caption{\textbf{Performance comparison in Simulation 1 and Simulation 3 (n = 1000).}
The blue lines represent estimated dose-response curves and corresponding 95\% confidence intervals (CIs) using cross-validated HAL (CV-HAL), while the orange lines represent those using undersmoothed HAL. Solid lines denote the oracle CIs derived from 500 Monte Carlo repetitions, and dashed lines indicate delta-method-based CIs. For the simpler curve in Simulation 1, both CV-HAL and undersmoothed HAL perform similarly. In contrast, for the more complex curve in Simulation 3, undersmoothing improves performance by yielding higher CI coverage rates and lower mean squared error (MSE). Additionally, delta-method-based CIs closely approximate the oracle CIs, demonstrating their accuracy.}
\label{fig:CV_vs_U_simu13}
\end{figure}

Table \ref{tab:CV_vs_U} summarizes the average pointwise performance of estimators at two points on the dose-response curve, using first-order Highly Adaptive Lasso (HAL) with $\lambda$ selected by both Cross-Validation (CV) and Undersmoothing criteria, for a sample size of 1000. In simpler scenarios, such as simulation 1, differences between the two selection methods were minimal. The delta-method-based confidence intervals achieved average coverage rates of 94.2\% and 95.7\% for CV and undersmoothing selectors, respectively. Both methods show an average bias of approximately 0.02. Figure~\ref{fig:CV_vs_U_simu13} illustrates that the CV selector slightly outperforms the undersmoothing selector, though the difference is negligible.

In more complex settings, such as simulation 3, the undersmoothing selector significantly outperformed the CV selector, especially at points of increased complexity along the curve. The CV selector yielded an oracle coverage rate of $83.1\%$ with an average bias of 0.048, while the Undersmoothing selector achieved a $90.4\%$ oracle coverage rate with a lower bias of 0.044. As shown in Figure~\ref{fig:CV_vs_U_simu13} and Table~\ref{tab:CV_vs_U}, the undersmoothed HAL estimators reduced bias and improved confidence interval coverage at points of increased complexity.

Furthermore, delta-method-based variance estimates closely approximated oracle variance in all simulations, as demonstrated in Figure~\ref{fig:CV_vs_U_simu13}. The dashed lines represent the pointwise performance of delta-method-based estimators and their inference, while the solid lines denote the oracle performance, highlighting the consistency between the two.

\begin{table}
\caption{\textbf{Average pointwise performance of the HAL-based plug-in estimator at $A=1.0$ and $A=4.2$ across four simulation settings (n = 1000).} The table compares the performance of the HAL estimator under two selection strategies: cross-validation (CV) and undersmoothing (U). For each simulation scenario and evaluation point, we report bias, mean squared error (MSE), confidence interval (CI) coverage, bias-to-standard error ratio (Bias/SE), and standard error (SE), using both delta-method-based and oracle CI constructions. In general, undersmoothing improves CI coverage and yields smaller Bias/SE ratios, particularly in more complex simulation settings (Simulations 3 and 4). However, this comes with a slight trade-off in increased standard error. The oracle results serve as a benchmark for evaluating the accuracy of the delta-method-based inference.}
\label{tab:CV_vs_U}
\begin{center}
\begin{tabular}{lllllllllll}
\hline
Simu &  &     &    &    & \multicolumn{3}{l}{Delta-method}  & \multicolumn{3}{l}{Oracle} \\ 
   & A& Selector & $|$Bias$|$ & MSE  & CI cov. & Bias/SE & SE  & CI cov. & Bias/SE & SE   \\ \hline
1  & 1.0 & CV    & 0.019 & 0.001 & 0.940    & 0.835  & 0.024 & 0.954    & 0.851  & 0.022 \\
   &     & U     & 0.021 & 0.001 & 0.974    & 0.689  & 0.031 & 0.954    & 0.795  & 0.026 \\
   & 4.2 & CV    & 0.020 & 0.001 & 0.944    & 0.830  & 0.025 & 0.954    & 0.785  & 0.025 \\
   &     & U     & 0.026 & 0.001 & 0.946    & 0.813  & 0.033 & 0.946    & 0.770  & 0.034 \\ \hline
2  & 1.0 & CV    & 0.027 & 0.001 & 0.834    & 1.095  & 0.027 & 0.912    & 0.924  & 0.029 \\
   &     & U     & 0.027 & 0.001 & 0.878    & 0.964  & 0.032 & 0.943    & 0.797  & 0.034 \\
   & 4.2 & CV    & 0.027 & 0.001 & 0.530    & 2.065  & 0.014 & 0.888    & 1.057  & 0.025 \\
   &     & U     & 0.022 & 0.001 & 0.669    & 1.739  & 0.015 & 0.931    & 0.880  & 0.025 \\ \hline
3  & 1.0 & CV    & 0.021 & 0.001 & 0.958    & 0.807  & 0.028 & 0.944    & 0.803  & 0.026 \\
   &     & U     & 0.025 & 0.001 & 0.964    & 0.719  & 0.037 & 0.946    & 0.781  & 0.032 \\
   & 4.2 & CV    & 0.062 & 0.005 & 0.614    & 1.817  & 0.037 & 0.768    & 1.323  & 0.047 \\
   &     & U     & 0.055 & 0.004 & 0.728    & 1.476  & 0.041 & 0.862    & 1.094  & 0.050 \\\hline
4  & 1.0 & CV    & 0.025 & 0.001 & 0.960    & 0.769  & 0.033 & 0.938    & 0.826  & 0.030 \\
   &     & U     & 0.028 & 0.001 & 0.958    & 0.692  & 0.044 & 0.948    & 0.789  & 0.035 \\
   & 4.2 & CV    & 0.112 & 0.017 & 0.482    & 2.134  & 0.052 & 0.636    & 1.643  & 0.068 \\
   &     & U     & 0.071 & 0.009 & 0.758    & 1.313  & 0.052 & 0.870    & 0.957  & 0.075 \\ \hline
\end{tabular}
\end{center}
\end{table}

\subsection{Evaluation of Undersmoothing Criteria}

\begin{figure}[h]
\begin{center}
    \includegraphics[width=0.8\textwidth]{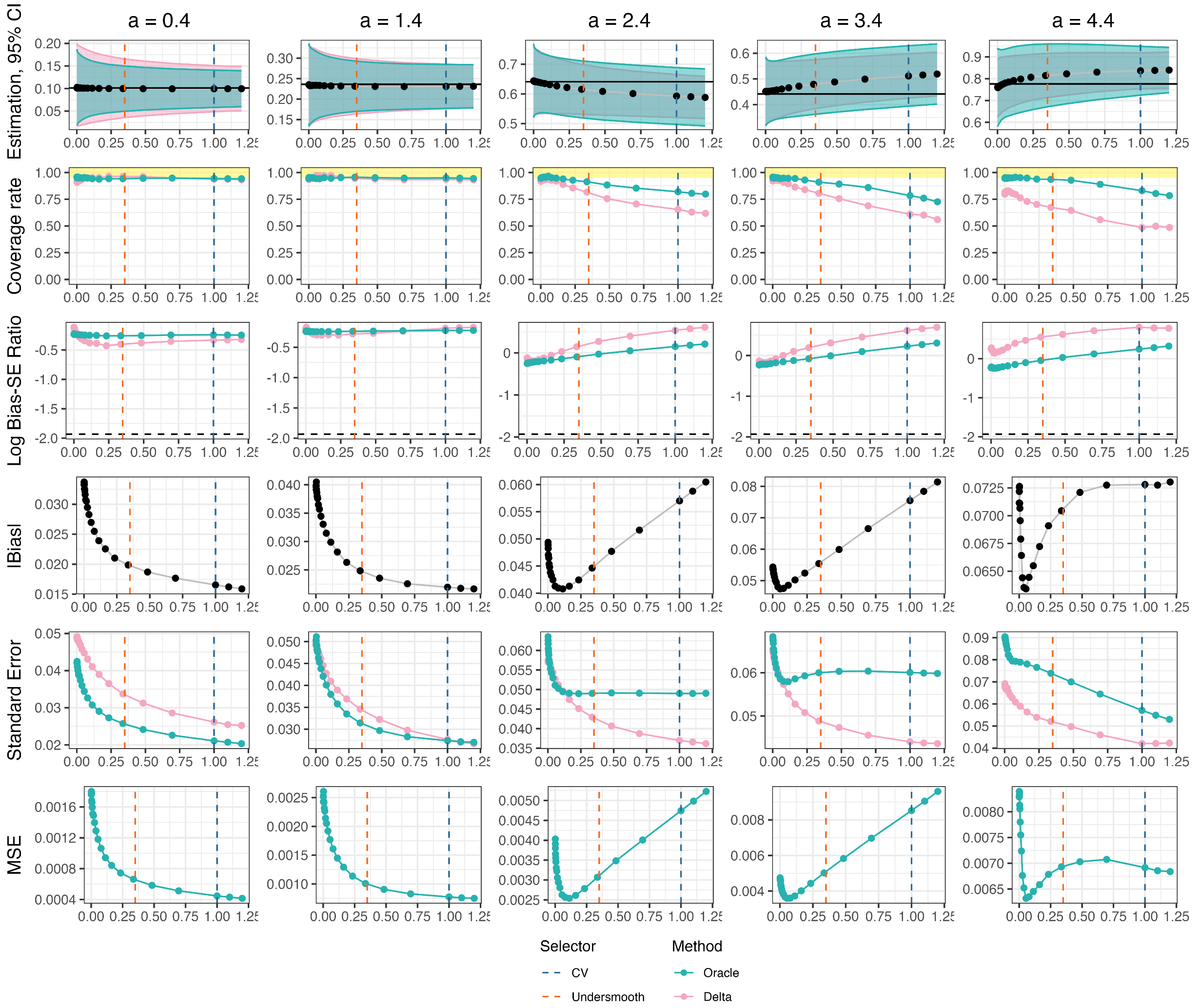}
\end{center}
\caption{\textbf{Pointwise performance across a grid of penalty values at five example locations in Simulation 3 (n = 1000).} Each column corresponds to a target point along the dose-response curve, with increasing complexity from left to right. The rows display: (1) estimated values and 95\% CIs; (2) CI coverage rates; (3) log bias-to-standard error ratio; (4) absolute bias; (5) standard error; and (6) mean squared error (MSE). Results aggregated over 500 Monte Carlo repetitions are shown in green (oracle-based) and pink (delta-method-based). The dashed blue vertical lines indicate the penalty selected by cross-validation (CV), and the dashed orange lines indicate the penalty selected by the undersmoothing criterion. The undersmoothing approach consistently selects smaller penalties and often strikes a better balance across bias, variance, and MSE, particularly in regions with higher curvature.}
\label{fig:grid_scalers}
\end{figure}

Figure \ref{fig:grid_scalers} illustrates the estimator's performance across a range of penalty candidates, delineated at five points along the dose-response curve in simulation 3, with a sample size of 1000. Each subplot corresponds to a different performance metric and treatment value $a$, showing how performance changes as the penalty values are adjusted. The orange vertical dashed lines represent penalties selected by the undersmoothing criteria, while the blue vertical dashed lines represent penalties selected by cross-validation (CV).

As penalty values decrease, we observe an increase in standard errors (SE) and coverage rates of 95\% confidence intervals (CIs). At certain points along the curve, such as $a = 0.4$ and $a = 1.4$, there is a monotonic increase in both bias and mean squared error (MSE) as penalties exceed the CV-selected value. In other cases, bias and MSE decrease initially, reaching a minimum before increasing again as the penalty further decreases.

Across the various treatment points ($a$ values), the undersmoothing selector achieves balanced performance, neither excessively penalizing nor underfitting. This balance is evidenced by the convergence of bias, standard error, and MSE toward optimal values at the selected penalties.

\subsection{Analysis of the Smoothness-Order Adaptive HAL}
In simulation 1, across all sample sizes, the first-order smoothness estimators consistently outperformed those with other smoothness orders, as shown in Appendix Figure~\ref{fig:smooth_order} and~\ref{fig:cv_risk_adapt}. The adaptability of the method is evident, with the Smoothness-Order Adaptive HAL selecting first-order smoothness in 70.6\% of the 500 Monte Carlo iterations for a sample size of 200, and in 98.4\% of the iterations for a sample size of 5000.

In simulation 3, interesting dynamics emerge. For smaller sample sizes (200, 500, and 1000), zero-order HAL slightly outperformed first-order HAL on the dose-response curve. However, the average CV risks were comparable or marginally higher for zero-order HAL relative to first-order HAL. 
As the sample size increased, first-order smoothness estimators consistently outperformed all other smoothness orders across the evaluation metrics. Furthermore, the smoothness level selected by the discrete Super Learner adapted to the sample size of the training data.
For smaller sample sizes of 200, 500, and 1000, the method based upon adaptive selection of spline order  chose first-order smoothness 56\%, 62.1\%, and 67.4\% of the time, respectively. For the largest sample size of 5000, the HAL demonstrated a definitive preference, selecting first-order smoothness in all iterations (100\%).
Such adaptive choices of smoothness order consistently align with the smoothness order that performed best for the available data.

\subsection{Comparison with Other Methods}

\begin{figure}
\begin{center}\includegraphics[width=0.8\textwidth]{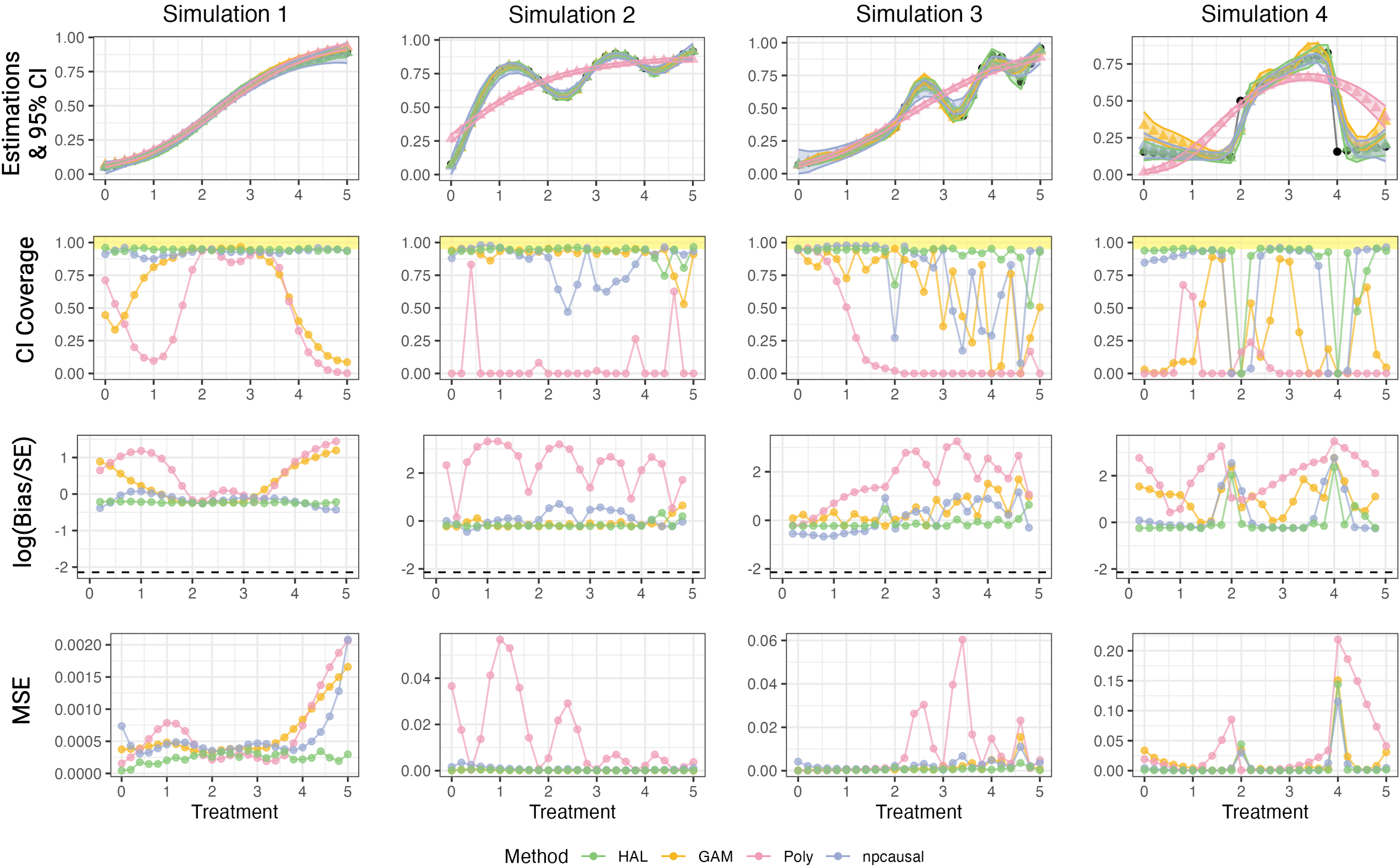} 
\end{center}
\caption{\textbf{Oracle performance of four estimators across four simulation settings (n = 5000).} Each column corresponds to a simulation scenario with increasing complexity from left to right. The rows display: (1) point estimates and 95\% confidence intervals (CIs), (2) CI coverage rates, (3) log bias-to-standard error (SE) ratio, and (4) mean squared error (MSE). The four estimators include HAL (green), GAM (orange), polynomial regression (pink), and nonparametric causal regression (npcausal, purple). In Simulation 1, all methods approximate the true curve well, with HAL and npcausal slightly outperforming the others. In Simulations 2–4, polynomial regression fails to capture the complex curve shapes, while HAL consistently demonstrates superior performance across all metrics, including accuracy, CI coverage, and MSE.}
\label{fig:compare}
\end{figure}

\begin{table}
\caption{\textbf{Average performance of four estimators across four simulation settings (n = 5000).}
This table summarizes the average bias, mean squared error (MSE), confidence interval (CI) coverage, bias-to-standard error ratio (Bias/SE), and standard error (SE) for four methods: Highly Adaptive Lasso (HAL), nonparametric causal regression (npcausal), generalized additive models (GAM), and polynomial regression (Poly). Results are reported for both oracle-based and delta-method-based inference procedures. HAL consistently achieves the best overall performance, with low bias, high CI coverage, and balanced Bias/SE across all simulation scenarios. While npcausal performs competitively in some settings, both GAM and polynomial regression exhibit deteriorating performance as the complexity of the data-generating process increases, particularly in Simulations 3 and 4. Polynomial regression fails to maintain accurate CI coverage and has the highest Bias/SE ratios and MSEs under complex scenarios.}
\label{tab:compare}
\begin{center}
\begin{tabular}{llllllllll}
\hline
Simu        &     &    &    & \multicolumn{3}{l}{Oracle}  & \multicolumn{3}{l}{Delta-method} \\ 
          & Method  & $|$Bias$|$ & MSE  & CI cov. & Bias/SE & SE  & CI cov. & Bias/SE & SE   \\ \hline 
1         & \textbf{HAL}   & \textbf{0.012} & \textbf{0.000} & \textbf{0.946}  & \textbf{0.792}  & \textbf{0.015} & \textbf{0.969}    & \textbf{0.710}  & 0.017 \\
          & npcausal & 0.014 & 0.001 & 0.927    & 0.841  & 0.017 & 0.868    & 1.065  & 0.014 \\
          & GAM   & 0.019 & 0.001 & 0.653    & 1.621  & 0.013 & 0.704    & 1.524  & 0.015 \\
          & Poly   & 0.021 & 0.001 & 0.498    & 2.153  & 0.010 & 0.593    & 1.848  & \textbf{0.012} \\ \hline
2         & \textbf{HAL}   & 0.010 & \textbf{0.000} & \textbf{0.930}    & \textbf{0.860}  & 0.012 & \textbf{0.889}    & \textbf{0.958}  & 0.012 \\
          & npcausal & 0.018 & 0.001 & 0.850    & 1.112  & 0.018 & 0.706    & 1.499  & 0.013 \\
          & GAM   & \textbf{0.009} & \textbf{0.000} & 0.907    & 0.919  & 0.010 & 0.876    & 0.998  & 0.009 \\
          & Poly   & 0.100 & 0.015 & 0.070    & 12.769 & \textbf{0.008} & 0.065    & 14.020  & \textbf{0.007} \\ \hline
3         & \textbf{HAL}   & \textbf{0.019} & \textbf{0.001} & \textbf{0.909}    & \textbf{0.916}  & 0.020 & \textbf{0.903}    & \textbf{0.898}  & 0.022 \\
          & npcausal & 0.034 & 0.003 & 0.764    & 1.260  & 0.028 & 0.614    & 1.900  & 0.020 \\
          & GAM   & 0.031 & 0.002 & 0.658    & 1.745  & 0.017 & 0.672    & 1.712  & 0.017 \\
          & Poly   & 0.073 & 0.010 & 0.214    & 7.528  & \textbf{0.009} & 0.253    & 6.068  & \textbf{0.011} \\ \hline
4         & \textbf{HAL}   & \textbf{0.042} & \textbf{0.008} & \textbf{0.843}    & \textbf{1.522}  & 0.024 & \textbf{0.840}    & \textbf{1.483}  & 0.027 \\
          & npcausal & 0.048 & \textbf{0.008} & 0.708    & 2.490  & 0.021 & 0.662    & 2.923  & 0.017 \\
          & GAM   & 0.087 & 0.014 & 0.281    & 3.807  & 0.022 & 0.338    & 3.243  & 0.025 \\
          & Poly   & 0.158 & 0.042 & 0.072    & 11.135 & \textbf{0.015} & 0.096    & 11.458  & \textbf{0.015} \\ \hline
\end{tabular}
\end{center}
\end{table}

Figure \ref{fig:compare} presents the oracle pointwise performance of four estimators in estimating the dose-response curve across four different simulations with a sample size of 5000. The estimators include the undersmoothed Highly Adaptive Lasso-based plug-in estimator (HAL), the Generalized Additive Model-based plug-in estimator (GAM), the polynomial model-based estimator (Poly), and the nonparametric causal inference estimator (npcausal). In addition, Table \ref{tab:compare} summarizes the average performance of the four estimators based on the oracle SE and delta method, reporting absolute bias, mean squared error (MSE), confidence interval (CI) coverage, bias-to-SE ratio, and SE across all simulations.

The first row of Figure \ref{fig:compare} shows the estimated dose-response curves and corresponding oracle confidence intervals (CIs). HAL, GAM, and npcausal estimators closely follow the trend of the true dose-response curves across all simulations, whereas the polynomial estimator demonstrates noticeable misalignment. The second row of Figure~\ref{fig:compare} shows the CI coverage rates for each estimator. In Simulation 1, the HAL estimator achieved an average oracle CI coverage rate of 94.6\% and a delta-method-based CI coverage rate of 96.9\%. In contrast, the other methods were more anti-conservative: npcausal had an average oracle CI coverage rate of 92.7\% and a delta-method CI coverage rate of 86.8\%, with GAM and polynomial regression performing even worse in terms of CI coverage.

Performance varied between simulations with differing complexities. In the more complex Simulation 3, the HAL estimator maintained strong performance with average oracle and delta-method CI coverage rates of 90.9\% and 90.3\%, respectively. The other methods, however, showed more pronounced anti-conservatism in this setting.

Across all simulations, the HAL estimator consistently demonstrated lower average absolute bias, with values of 0.012, 0.010, 0.019, and 0.042. While the npcausal estimator exhibited slightly higher absolute bias, it still outperformed the GAM and polynomial estimators. For both MSE and bias-to-SE ratio, HAL estimators had superior results, consistently achieving lower variance and bias ratios, as detailed in Table \ref{tab:compare}.

\section{Discussions}
\label{s:discussion}
In this paper, we explored the performance of different methods for causal dose-response estimation through a series of simulations, with a novel emphasis on the Highly Adaptive Lasso (HAL) plug-in estimator. Our contribution lies in implementing HAL with inference in the context of dose-response estimation and systematically evaluating its empirical robustness and adaptability across a range of complex scenarios, areas which have not been thoroughly addressed in previous work. We investigated the reliability of the HAL estimator and examined estimation criteria, including penalty selection and smoothness order choices.

Our results highlight the importance the of choice between cross-validation (CV) and undersmoothing when selecting HAL's penalty parameter. In simple scenarios, such as in simulation 1, where the dose-response curve is smooth, CV and undersmoothing approaches exhibit comparable performances. However, as curve complexity increases, such as in simulation 3, the undersmoothing approach stands out by achieving higher oracle coverage rates and lower biases. This finding indicates that undersmoothing is particularly beneficial for accurately capturing complex dose-response relationships, striking an optimal balance between bias and variance. Importantly, undersmoothing enables HAL to adjust across various treatment levels on the curve effectively. Additionally, smoothness-adaptive HAL selects the optimal smoothness orders given the dose-response curve's complexity and available sample size. With sufficiently large sample sizes, first-order HAL demonstrates the flexibility to handle both simple and complex (non-smooth) curve structures.

When comparing the four estimators—HAL, Generalized Additive Models (GAM), polynomial regression, and the nonparametric causal inference approach (npcausal)—our results show that the HAL-based plug-in estimator consistently delivers superior performance. It achieves the highest CI coverage rates, lowest absolute biases, and smallest mean squared error (MSE) values, especially in scenarios with more complex dose-response curves. In contrast, other estimators, particularly the polynomial model, show considerable variability and fail to maintain the same level of performance. These findings establish the HAL-based plug-in estimator as a preferred method for dose-response estimation, due to its robustness, accuracy, and ability to capture underlying relationships without relying on restrictive parametric assumptions.

The estimator's ability to provide unbiased estimates of complex dose-response relationships and deliver confidence intervals achieving nominal coverage (valid inference) further enhances its robustness and practical applicability. Its adaptability to increasing complexity and varying sample sizes, facilitated by smoothness-order adjustments, highlights its potential for broad application in diverse practical contexts.

While HAL shows considerable promise in modeling dose-response curves, there are computational challenges. Specifically, its implementation demands substantial computing resources, both in terms of time and memory, as it incorporates a comprehensive set of basis functions to flexibly fit a wide range of true curves without assuming specific parametric forms. Moreover, solving the inverse of a matrix with many basis functions, especially those selected with the undersmoothing criteria, may lead to invertibility issues. Ongoing research aims to address these concerns by improving HAL’s scalability and computational efficiency.

Our approach presents HAL’s capability of providing statistical inference for non-pathwise differentiable parameters, as demonstrated for the conditional mean and the conditional average treatment effect (Zhang et al., 2025; Zhang et al., 2024). A brief overview of this capability is also provided in Butzin‑Dozier et al. (2024). Moreover, advances in the adaptive targeted maximum likelihood framework (van~der~Laan et al., 2024) may further improve robustness for nonpathwise differentiable functions. These developments, coupled with the adoption of high-dimensional parametric working models revealed by an initial HAL fit, suggest substantial future improvements in estimation and inference for complex causal parameters.

\section{Conclusion}
In this paper, we presented the HAL-based plug-in estimator for continuous causal dose-response curves, highlighting both its theoretical underpinnings and empirical performance. By leveraging the flexibility of càdlàg functions and the adaptivity of HAL, the estimator achieves pointwise asymptotic normality under weak conditions, even for parameters that are non-pathwise differentiable.. Through extensive simulations, we demonstrated its robustness across diverse settings, consistently yielding valid coverage, reduced bias, and superior accuracy relative to established alternatives such as GAM, polynomial regression, and npcausal estimators. Importantly, undersmoothing and smoothness-adaptive selection further enhance performance, enabling the method to adapt to increasing complexity with growing sample sizes.

Together, these findings position the HAL plug-in estimator as a practical, theoretically grounded, and empirically validated approach for dose-response estimation in continuous treatment settings, bridging the gap between asymptotic theory and applied causal inference.

The halDRC R package, which includes both real-data application examples and simulated sample demos, is available at \url{https://github.com/SeraphinaShi/halDRC}. The source code implementing our methods can be found at \url{https://github.com/SeraphinaShi/HAL-DoseResponseCurve}. These resources support reproducibility and are intended to facilitate further research in this area.

\section*{Appendix}
\subsection*{A. Data Generating Distributions (DGDs) in 4 Simulations}

\begin{table}[h!]
\caption{List of Underling Data Generating Distributions (DGDs)}
\vspace{1ex}
\centering
\renewcommand{\arraystretch}{1.5} 
\begin{tabular}{|c|l|}
\hline
Simulation & DGD \\
\hline
1 & 
\begin{minipage}[t]{0.8\columnwidth}
\begin{itemize}
    \item Exogenous variables:
    \begin{itemize}
        \item $U_W \sim Normal(\mu=0, \sigma^2 = 1^2)$
        \item $U_A \sim Normal(\mu=0, \sigma^2 = 2^2)$
        \item $U_Y \sim Uniform(min = 0, max = 1)$
    \end{itemize}
    \item Structural equations F and endogenous variables:
    \begin{itemize}
        \item $W =  U_W$
        \item $A = bound(2 - 0.5W + U_A, min=0, max=5)$
        \item $Y = \mathbf{I}[U_Y < expit(-3 + 0.5W + 1.25A -0.5WA)]$
    \end{itemize}
    \vspace{1ex}
\end{itemize}
\end{minipage} \\
\hline
2 & 
\begin{minipage}[t]{0.8\columnwidth}
\begin{itemize}
    \item Exogenous variables:
    \begin{itemize}
        \item $U_W \sim Normal(\mu=0, \sigma^2 = 1^2)$
        \item $U_A \sim Normal(\mu=0, \sigma^2 = 1.3    ^2)$
        \item $U_Y \sim Uniform(min = 0, max = 1)$
    \end{itemize}
    \item Structural equations F and endogenous variables:
    \begin{itemize}
        \item $W =  U_W$
        \item $A = bound(2.5 - 0.5W + U_A, min=0, max=5)$
        \item $Y = \mathbf{I}[U_Y < expit(-5 + 3W + 5sin(1.25A^{1.5}) + 3WA)]$
    \end{itemize}
    \vspace{1ex}
\end{itemize}
\end{minipage} \\
\hline
3 & 
\begin{minipage}[t]{0.8\columnwidth}
\begin{itemize}
    \item Exogenous variables:
    \begin{itemize}
        \item $U_W \sim Normal(\mu=0, \sigma^2 = 1^2)$
        \item $U_A \sim Normal(\mu=0, \sigma^2 = 2^2)$
        \item $U_Y \sim Uniform(min = 0, max = 1)$
    \end{itemize}
    \item Structural equations F and endogenous variables:
    \begin{itemize}
        \item $W =  U_W$
        \item $A = bound(2 - 0.5W + U_A, min=0, max=5)$
        \item $Y = \mathbf{I}[U_Y < expit(-4 - 2W + 1.5A + \mathbf{I}(A>2) 1.5 sin((0.8A)^2 - 2.56)  )]$
    \end{itemize}
    \vspace{1ex}
\end{itemize}
\end{minipage} \\
\hline
4 & 
\begin{minipage}[t]{0.8\columnwidth}
\begin{itemize}
    \item Exogenous variables:
    \begin{itemize}
        \item $U_W \sim Normal(\mu=0, \sigma^2 = 1^2)$
        \item $U_A \sim Normal(\mu=0, \sigma^2 = 1^2)$
        \item $U_Y \sim Uniform(min = 0, max = 1)$
    \end{itemize}
    \item Structural equations and endogenous variables:
    \begin{itemize}
        \item $W = U_W$
        \item $A = bound(2.5 - 0.5W + U_A, min=0, max=5)$
        \item $Y = \mathbf{I}[U_Y < expit(-2 + W + A\mathbf{I}(A\geq2) - A\mathbf{I}(A\geq4) - 0.5WA)]$
    \end{itemize}
    \vspace{1ex}
\end{itemize}
\end{minipage} \\
\hline
\end{tabular}
\label{tab:DGDs}
\end{table}

\subsection*{B. Compare performances of estimators based on HAL with different smoothness order splines}

\begin{figure}[h]
\begin{center}
    \includegraphics[width=0.9\textwidth]{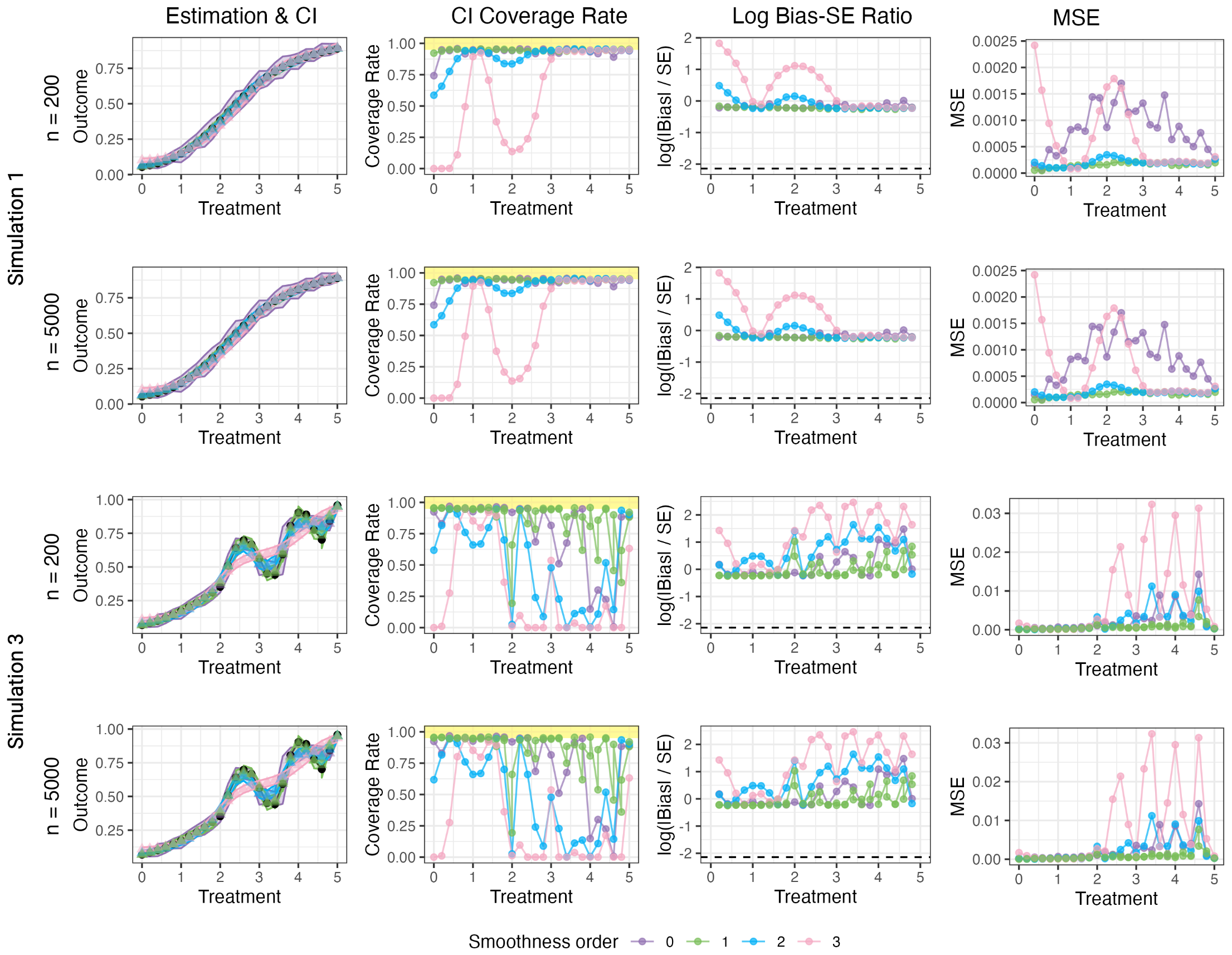} 
\end{center}
\caption{Oracle performance of estimators based on HAL trained with different smoothness orders across simulations 1 and 3 with sample sizes of 200 and 5,000}
\label{fig:smooth_order}
\end{figure}

\begin{figure}[h]
\begin{center}
    \includegraphics[width=0.9\textwidth]{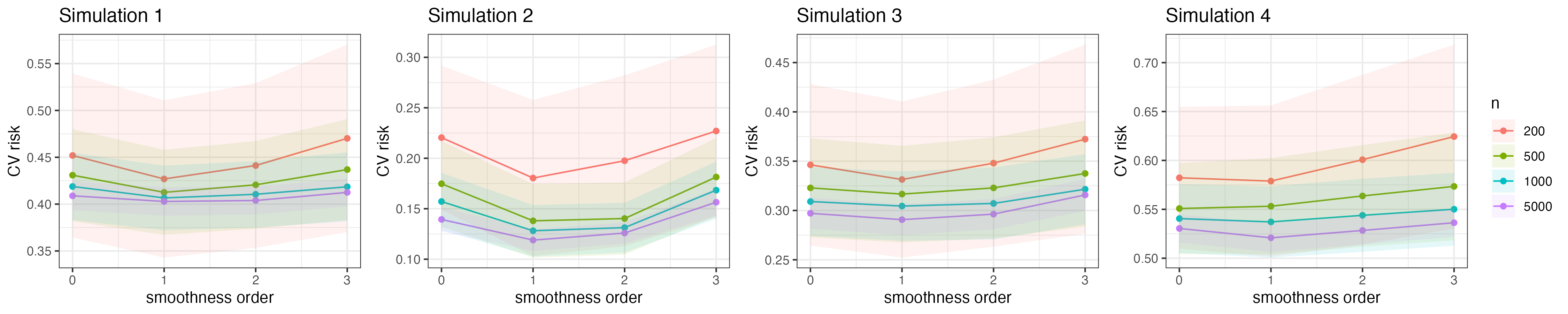} 
\end{center}
\caption{Average CV risks of estimators based on HAL trained with different smoothness orders across 4 simulations with all tested sample sizes}
\label{fig:cv_risk_adapt}
\end{figure}


\begin{thebibliography}{99}
\bibitem{van2015generally} van der Laan, M. J. (2015). A generally efficient targeted minimum loss based estimator. bepress.

\bibitem{van2017generally} van der Laan, M. (2017). A generally efficient targeted minimum loss based estimator based on the highly adaptive lasso. \textit{The International Journal of Biostatistics}, \textbf{13}(2), De Gruyter.

\bibitem{van2023higher} van der Laan, M. (2023). Higher Order Spline Highly Adaptive Lasso Estimators of Functional Parameters: Pointwise Asymptotic Normality and Uniform Convergence Rates. \textit{arXiv preprint arXiv:2301.13354}.

\bibitem{van2008direct} van der Laan, M. J., \& Petersen, M. L. (2008). Direct effect models. \textit{The International Journal of Biostatistics}, \textbf{4}(1), De Gruyter.

\bibitem{petersen2006estimation} Petersen, M. L., Sinisi, S. E., \& van der Laan, M. J. (2006). Estimation of direct causal effects. \textit{Epidemiology}, 276--284, JSTOR.

\bibitem{rubin2004direct} Rubin, D. B. (2004). Direct and indirect causal effects via potential outcomes. \textit{Scandinavian Journal of Statistics}, \textbf{31}(2), 161--170, Wiley Online Library.

\bibitem{van2011targeted} van der Laan, M. J., \& Rose, S. (2011). \textit{Targeted learning: Causal inference for observational and experimental data}. New York: Springer.

\bibitem{rubin2006matched} Rubin, D. B. (2006). \textit{Matched sampling for causal effects}. Cambridge University Press.

\bibitem{laan2003unified} van der Laan, M. J., \& Robins, J. M. (2003). \textit{Unified methods for censored longitudinal data and causality}. Springer.

\bibitem{abadie2006large} Abadie, A., \& Imbens, G. W. (2006). Large sample properties of matching estimators for average treatment effects. \textit{Econometrica}, \textbf{74}(1), 235--267, Wiley Online Library.

\bibitem{angrist1995identification} Angrist, J., \& Imbens, G. (1995). Identification and estimation of local average treatment effects. National Bureau of Economic Research, Cambridge, Mass., USA.

\bibitem{robins1986new} Robins, J. (1986). A new approach to causal inference in mortality studies with a sustained exposure period—application to control of the healthy worker survivor effect. \textit{Mathematical Modelling}, \textbf{7}(9-12), 1393--1512, Elsevier.

\bibitem{rubin1978bayesian} Rubin, D. B. (1978). Bayesian inference for causal effects: The role of randomization. \textit{The Annals of Statistics}, 34--58, JSTOR.

\bibitem{abadie2011bias} Abadie, A., \& Imbens, G. W. (2011). Bias-corrected matching estimators for average treatment effects. \textit{Journal of Business \& Economic Statistics}, \textbf{29}(1), 1--11, Taylor \& Francis.

\bibitem{austin2015moving} Austin, P. C., \& Stuart, E. A. (2015). Moving towards best practice when using inverse probability of treatment weighting (IPTW) using the propensity score to estimate causal treatment effects in observational studies. \textit{Statistics in Medicine}, \textbf{34}(28), 3661--3679, Wiley Online Library.

\bibitem{luque2018targeted} Luque-Fernandez, M. A., Schomaker, M., Rachet, B., \& Schnitzer, M. E. (2018). Targeted maximum likelihood estimation for a binary treatment: A tutorial. \textit{Statistics in Medicine}, \textbf{37}(16), 2530--2546, Wiley Online Library.

\bibitem{van2022efficient} van der Laan, M. J., Benkeser, D., \& Cai, W. (2022). Efficient estimation of pathwise differentiable target parameters with the undersmoothed highly adaptive lasso. \textit{The International Journal of Biostatistics}, De Gruyter.

\bibitem{bickel1993efficient} Bickel, P. J., Klaassen, C. A. J., Ritov, Y., \& Wellner, J. A. (1993). \textit{Efficient and adaptive estimation for semiparametric models}. Springer.

\bibitem{robins2000marginal} Robins, J. M., Hernan, M. A., \& Brumback, B. (2000). Marginal structural models and causal inference in epidemiology. \textit{Epidemiology}, 550--560, JSTOR.

\bibitem{neugebauer2007nonparametric} Neugebauer, R., \& van der Laan, M. (2007). Nonparametric causal effects based on marginal structural models. \textit{Journal of Statistical Planning and Inference}, \textbf{137}(2), 419--434, Elsevier.

\bibitem{haight2010cross} Haight, T. J., Wang, Y., van der Laan, M. J., \& Tager, I. B. (2010). A cross-validation deletion--substitution--addition model selection algorithm: Application to marginal structural models. \textit{Computational Statistics \& Data Analysis}, \textbf{54}(12), 3080--3094, Elsevier.

\bibitem{robins2000causal} Robins, J. M., \& Greenland, S. (2000). Causal inference without counterfactuals: Comment. \textit{Journal of the American Statistical Association}, \textbf{95}(450), 431--435, JSTOR.

\bibitem{hirano2004propensity} Hirano, K., \& Imbens, G. W. (2004). The propensity score with continuous treatments. \textit{Applied Bayesian Modeling and Causal Inference from Incomplete-Data Perspectives}, \textbf{226164}, 73--84.

\bibitem{callaway2021difference} Callaway, B., Goodman-Bacon, A., \& Sant'Anna, P. H. C. (2021). Difference-in-differences with a continuous treatment. \textit{arXiv preprint arXiv:2107.02637}.

\bibitem{schwab2020learning} Schwab, P., Linhardt, L., Bauer, S., Buhmann, J. M., \& Karlen, W. (2020). Learning counterfactual representations for estimating individual dose-response curves. \textit{Proceedings of the AAAI Conference on Artificial Intelligence}, \textbf{34}(04), 5612--5619.

\bibitem{diaz2013targeted} Díaz, I., \& van der Laan, M. J. (2013). Targeted data adaptive estimation of the causal dose-response curve. \textit{Journal of Causal Inference}, \textbf{1}(2), 171--192, De Gruyter.

\bibitem{van2018targeted} van der Laan, M. J., \& Rose, S. (2018). \textit{Targeted learning in data science}. Springer.

\bibitem{kennedy2017non} Kennedy, E. H., Ma, Z., McHugh, M. D., \& Small, D. S. (2017). Non-parametric methods for doubly robust estimation of continuous treatment effects. \textit{Journal of the Royal Statistical Society: Series B}, \textbf{79}(4), 1229--1245, Oxford University Press.

\bibitem{diaz2023nonparametric} 
Díaz, I., Williams, N., Hoffman, K. L., \& Schenck, E. J. (2023). 
Nonparametric causal effects based on longitudinal modified treatment policies. 
\textit{Journal of the American Statistical Association}, \textbf{118}(542), 846--857.


\bibitem{van1996weak} van der Vaart, A. W., \& Wellner, J. A. (1996). \textit{Weak convergence}. Springer.

\bibitem{van2007super} Van der Laan, M. J., Polley, E. C., and Hubbard, A. E. (2007). Super learner. \textit{Statistical Applications in Genetics and Molecular Biology}, 6(1). De Gruyter.

\bibitem{friedman2021package} Friedman, J., Hastie, T., Tibshirani, R., Narasimhan, B., Tay, K., Simon, N., \& Qian, J. (2021). Package ‘glmnet’. \textit{CRAN R Repositary}, \textbf{595}.

\bibitem{hejazi2020hal9001} Hejazi, N. S., Coyle, J. R., \& van der Laan, M. J. (2020). hal9001: Scalable highly adaptive lasso regression in R. \textit{Journal of Open Source Software}, \textbf{5}(53), 2526.

\bibitem{zhang2024evaluating} Zhang, W., Hudson, A., Petersen, M. and van der Laan, M. (2024). Evaluating and Utilizing Surrogate Outcomes in Covariate-Adjusted Response-Adaptive Designs. \textit{arXiv preprint arXiv:2408.02667}.

\bibitem{zhang2025constructing} Zhang, W., Shi, J., Hubbard, A. and van der Laan, M. (2025). Constructing Confidence Intervals for Infinite-Dimensional Functional Prameters by Highly Adaptive Lasso. \textit{arXiv preprint arXiv:2507.10511}.

\bibitem{butzin2024highly} Butzin-Dozier, Z., Qiu, S., Hubbard, A. E., Shi, J. S., \& van der Laan, M. J. (2024). Highly adaptive LASSO: Machine learning that provides valid nonparametric inference in realistic models. \textit{medRxiv}.

\bibitem{van2024adaptive} van der Laan, M., Qiu, S., and van der Laan, L. (2024). Adaptive-TMLE for the Average Treatment Effect based on Randomized Controlled Trial Augmented with Real-World Data. \textit{arXiv preprint arXiv:2405.07186}.
\end{thebibliography}
\end{document}